\documentclass[twocolumn,groupedaddress,floatfix,prb]{revtex4}
\usepackage{graphicx}

\begin{document}

\title{Probing phase coexistence and stabilization of the spin--ordered ferrimagnetic state by Calcium addition in the Y(Ba$_{1-x}$Ca$_{x}$)Co$_2$O$_{5.5}$
layered cobaltites using neutron diffraction}
\author{G. Aurelio}
\email[Corresponding Author: ]{gaurelio@cab.cnea.gov.ar}
 \altaffiliation[Also at ]{CIC, Consejo Nacional de Investigaciones Cient\'{\i }ficas y T\'{e}cnicas.}
\author{J. Curiale}  \altaffiliation[Also at ]{Instituto Balseiro, Universidad Nacional de Cuyo.}
\author{R. D. S\'{a}nchez}  \altaffiliation[Also at ]{CIC, Consejo Nacional de Investigaciones Cient\'{\i }ficas y T\'{e}cnicas.} \altaffiliation[Also at ]{Instituto Balseiro, Universidad Nacional de Cuyo.}

\affiliation{Comisi\'{o}n Nacional de Energ\'{\i}a At\'{o}mica -- Centro
At\'{o}mico Bariloche, Av. Bustillo 9500, 8400 S. C. de Bariloche, RN,
Argentina}

\author{G. J. Cuello}
\affiliation{Institut Laue Langevin, BP 156, F-38042 Grenoble Cedex
9, France}

\date{\today}

\begin{abstract}
In this article we study the effects of a partial substitution of Ba with the
smaller cation Ca in the layered cobaltites YBaCo$_2$O$_{5+\delta}$ for
$\delta \approx 0.5$. Neutron thermodiffractograms are reported for the
compounds YBa$_{0.95}$Ca$_{0.05}$Co$_2$O$_{5.5}$ ($x_{\rm{Ca}}=0.05$) and
YBa$_{0.90}$Ca$_{0.10}$Co$_2$O$_{5.5}$ ($x_{\rm{Ca}}=0.10$) in the temperature
range 20~K~$\leq T \leq$~300~K, as well as high resolution neutron diffraction
experiments at selected temperatures for the samples $x_{\rm{Ca}}=0.05$,
$x_{\rm{Ca}}=0.10$ and the parent compound $x_{\rm{Ca}}=0$. We have found the
magnetic properties to be strongly affected by the cationic substitution.
Although the ``122" perovskite structure seems unaffected by Ca addition, the
magnetic arrangements of Co ions are drastically modified: the
antiferromagnetic (AFM) long--range order is destroyed, and a ferrimagnetic
phase with spin state order is stabilized below $T \sim 290$~K. For the sample
with $x_{\rm{Ca}}=0.05$ a fraction of AFM phase coexists with the
ferrimagnetic one below $T \sim 190$~K, whereas for $x_{\rm{Ca}}=0.10$ the AFM
order is completely lost. The systematic refinement of the whole series has
allowed for a better understanding of the observed low--temperature
diffraction patterns of the parent compound, YBaCo$_2$O$_{5.5}$, which had not
yet been clarified. A two--phase scenario is proposed for the $x_{\rm{Ca}}=0$
compound which is compatible with the phase coexistence observed in the
$x_{\rm{Ca}}=0.05$ sample.

\end{abstract}

 \maketitle

\section{Introduction \label{Introduction}}

During the last decade, cobaltites have gained increased attention. A great
effort is being made to clarify and systematize the extremely rich variety of
phenomena they exhibit. Initially, they were expected to show similar
properties to other perovskite--family members, such as manganites and
cuprates,~\cite{97Mar,01Res,03Kha,05Tas,05Zho} but soon it was found that they
present additional tunable features, as the cobalt spin state, that add to
their complexity but also make them even more fascinating and challenging.
Among cobaltites, the layered compounds \emph{R}BaCo$_{2}$O$_{5+\delta}$
(\emph{R} being a rare earth) are currently being intensively
studied.~\cite{03Kha,05Tas,05Zho,05Fro} The oxygen content in these compounds
can be modified in a wide range ($0 \leq \delta \leq 0.9$ depending on the
\emph{R} cation and the synthesis conditions),~\cite{05Fro} which in turn
controls the mixed valence state of Co ions. Several factors strongly
influence the physical properties of these cobaltites: the non--stoichiometry,
the \emph{R} cation size, the vacancies structural order, and --- as we will
show in the present work--- also the structural disorder introduced by doping
the Ba site with small quantities of smaller cations with the same valence
state.

From a structural point of view, \emph{R}BaCo$_{2}$O$_{5+\delta}$ is formed by
a stacking sequence of [CoO$_2$]--[BaO]--[CoO$_2$]--[\emph{R}O$_{\delta}$]
planes along the $c-$axis,~\cite{00Mor,01Res} the usually called ``112"
structure derived form the $a_{P}$x$a_{P}$x$2a_{P}$ cell, being $a_{P}$ the
perovskite unit cell constant. The symmetry may be tetragonal or orthorhombic,
depending on the oxygen content and the \emph{R} cation. The oxygen vacancies
have a strong tendency to become ordered, which results in several
superstructures.~\cite{01Aka}

Of particular interest is the case $\delta=0.5$, for which Co is expected to
be completely in the +3 valence state. In this case, a particular order of
oxygen vacancies leads to the ``122" superstructure, consisting of an ordered
array of 50\% Co atoms in octahedral oxygen coordination and 50\% in a
pyramidal environment. This, in turn, favors a metal--insulator (MI)
transition just above room temperature (the $T_{\rm{MI}}$ depends again on the
\emph{R} cation) which can be found only for $\delta$ values very close to
0.5.~\cite{05Tas} When doping with holes (Co$^{4+}$, $\delta>0.5$) these
compounds behave as metals above the $T_{\rm{MI}}$ transition, but when doping
with electrons (Co$^{2+}$, $\delta<0.5$), these do not seem to participate in
charge transport, which has been explained in terms of a spin
blockade.~\cite{04Mai} There has arisen a big controversy regarding the
physical phenomena which occur at $T_{\rm{MI}}$. Regardless of the \emph{R}
cation, cobaltites with $\delta \sim 0.5$ all show a jump in resistivity and a
concomitant lattice distortion with a sudden volume collapse. The distortion
is associated with specific changes in the Co--O distances in pyramids and
octahedra, and became the subject of different interpretations. A possible
driving force for the MI transition has been proposed to be a spin state
transition from the Co low--spin state (LS: $t_{2g}^{6} e_{g}^{0}$) to a
high--spin state (HS: $t_{2g}^{4} e_{g}^{2}$) occurring only at the octahedral
sites.~\cite{02Fro} For the particular cases of \emph{R}~=~Pr~\cite{06Fro} and
Gd,~\cite{02Fro,03Fro} this hypothesis would also be supported by a change in
the slope of the inverse susceptibility curve at $T_{\rm{MI}}$, which has been
analyzed in terms of the Curie--Weiss model. Further support to this scenario
was given by Maignan \textit{et al.}~\cite{04Mai} for \emph{R}~=~Ho based on
thermoelectric measurements, showing that the spin blockade mechanism is fully
compatible with this picture. Other authors have proposed that a
$t_{2g}-e_{g}$ hybridization is enhanced in the metallic phase by the lattice
distortion, such that the metallic or insulating behavior would be determined
by the intersite mixing of the itinerant 3d electrons between the octahedral
and pyramidal sites.~\cite{06Tak} For \emph{R}~=~Tb the distortions of
pyramids and octahedra were interpreted as a
$d_{3x^{2}-r^{2}}/d_{3y^{2}-r^{2}}$ orbital ordering transition accompanied by
a intermediate--spin (IS: $t_{2g}^{5} e_{g}^{1}$)to HS spin state
transition.~\cite{00Mor} Furthermore, Pomjakushina \textit{et
al.}~\cite{06Pom} proposed that the observed volume collapse at the transition
temperature and the existence of an isotopic effect are indicative of a charge
delocalization breaking the orbital order of the insulating phase, which could
be compatible with a spin state switch, but occurring in pyramids, not in
octahedra. It seems obvious that this issue is far from being clarified and
some effort must be made to systematize the study of the MI transition in
cobaltites.

A second controversy, closely related to the one mentioned in the previous
paragraph, concerns the low temperature ordering of the magnetic moments at
the Co sites. Again, the feature which seems to be common to all \emph{R}
cobaltites is the existence of a spontaneous magnetization in a more or less
narrow temperature range, depending on \emph{R}, below room temperature. Above
this range they are paramagnetic, and below this range they transform to an
antiferromagnetic (AFM) state. It is now well established that the origin of
the spontaneous magnetization is not a ferromagnetic order, but among the two
remaining possibilities, \textit{i.e.}, a ferrimagnetic phase or a canted AFM
phase, there are various different models which have been proposed. Some of
these models involve a so--called spin state ordering (SSO) in which not only
the spin state may be different between Co atoms located at pyramids and Co
atoms located at octahedra, but also among the pyramidal~\cite{05Pla} or the
octahedral sites~\cite{02Fau,05Kha} a SSO may arise leading to a doubling of
the $a-$axis in the unit cell. Indeed, a theoretical work by Khomskii and
L\"{o}w~\cite{04Kho} showed that such spin superstructures can be
energetically favorable. These models would correspond to a ferrimagnetic
phase. On the other hand, the proposers of canted--AFM models argue that there
are no structural evidences for the doubling of the $a-$axis, and adopt the
canted models which also explain the neutron diffraction data, with a doubling
of the $a-$axis just in the magnetic cell.~\cite{03Sod,06Fro} However, some
care must be taken when comparing all these experimental data. For instance,
there may be no evidence of a ``222" superstructure in cobaltites with
\emph{R}~=~Gd and Pr~\cite{06Fro} but the case might be different for other
lanthanides. In fact, some studies using NMR techniques showed that for
\emph{R}~=~Y, there are four non--equivalent Co sites at low
temperature,~\cite{03Ito} and for \emph{R}~=~Eu there are three,~\cite{04Kub}
which is compatible with the SSO scenario. It should be emphasized, too, that
the ``222" superstructure is very hard to detect from diffraction measurements
unless an exceptionally high signal--to--noise ratio is attained. Using
transmission geometry, Chernenkov \textit{et al.}~\cite{05Che} have shown that
the superstructure can indeed be observed in single crystals with
\emph{R}~=~Gd using X--ray diffraction. In all cases, there seems to be
consensus on the IS character of pyramidal Co
atoms,~\cite{02Fro,05Zho,01Kus,01Pou,06Par} although the spin state ---or spin
states--- at octahedral sites remains uncertain or may, at least, depend on
the \emph{R} size.

Most studies of layered cobaltites were conducted for \emph{R} among the
lanthanides, but the compound with \emph{R}~=~Y$^{3+}$, which is a small,
non--magnetic ion, is a good candidate to isolate the intrinsic properties of
Co and explore the small--\emph{R} region of the phase diagram. It is now well
documented, for instance, that the MI transition temperature decreases with
the \emph{R} size. To gain more insight into the possible role of disorder, we
have introduced a second source of distortion, by substituting the Ba--site
with Ca, which has a smaller atomic radius. In addition, it has recently been
postulated on the basis of density--functional theory calculations, that a
smaller cation substitution in the Ba--site of small lanthanide cobaltites
could be a promising compound to exhibit enhanced giant magnetoresistance
properties.~\cite{01Wu} The present work is aimed at characterizing and
correlating the Ba--substituted compounds when compared to the parent
YBaCo$_{2}$O$_{5.5}$ cobaltite. We have performed a structural
characterization using neutron powder diffraction (NPD) to study the interplay
between the structures and their magnetic order, and correlate this
information with our previous magnetic studies.~\cite{law3m,07Aur}

\section{Experimental methods \label{experimental}}

Three polycrystalline samples were prepared by solid--state reaction.
High--purity powders of Y$_{2}$O$_{3}$, BaCO$_{3}$, CaCO$_{3}$ and
Co$_{3}$O$_{4}$ were mixed at stoichiometric weights to prepare the compounds
YBaCo$_{2}$O$_{5+ \delta}$ ($x_{\rm{Ca}}=0$),
YBa$_{0.95}$Ca$_{0.05}$Co$_2$O$_{5.5}$ ($x_{\rm{Ca}}=0.05$) and
YBa$_{0.90}$Ca$_{0.10}$Co$_2$O$_{5.5}$ ($x_{\rm{Ca}}=0.10$). After a
de--carbonation process at 1173~K for 18~h, the mixtures were pressed into
pellets and annealed. The samples were annealed together during 25~h at 1273~K
and slowly cooled at 1~K/min in oxygen flow. After a regrinding of the
resulting pellets, the compression and annealing at 1273~K in oxygen processes
were repeated. A single batch was used for all the samples to guarantee
identical synthesis conditions, which resulted in samples of about 1.5~g.

The oxygen content in our samples has been determined by refinement of our NPD
data. In addition, we have compared the macroscopic magnetization and
resistivity of our $x_{\rm{Ca}}=0$ sample with a very detailed study of the
parent compound YBaCo$_{2}$O$_{5+ \delta}$ early reported by Akahoshi and
Ueda~\cite{99Aka}. Their work presents the existing correlation between oxygen
content and magnetic and transport properties. In particular, the
magnetization curve for our sample (Fig.~\ref{f:MdeT}) reveals an excellent
quantitative agreement with their results for $\delta=0.5$, and a clear
disagreement outside the range $0.44 < \delta \leq 0.52$. Moreover, the
resistivity measurements in our samples~\cite{07Aur} show the characteristic
sharp jump of the MI transition, which has been shown to occur only for $0.45
< \delta \leq 0.65$ but only to be sharp for $\delta \simeq 0.5$~\cite{05Tas}.
These limits give us confidence in the refined values from our NDP data. The
global oxygen contents refined independently from 8 high resolution
diffractograms, corresponding to different samples and temperatures
---always below 350~K--- were in mutual agreement within experimental error, yielding
an average value of $\delta=0.46\pm 0.02$. In the following we shall refer to
the samples using the notation Y(Ba,Ca)Co$_2$O$_{5.5}$.

Neutron thermodiffraction data were collected on the high--intensity two--axis
diffractometer D20 located at the High Flux Reactor of Institute
Laue--Langevin ILL, Grenoble, France. Samples with $x_{\text{Ca}}=0.05$ and
$x_{\text{Ca}}=0.10$ were cooled in a standard orange cryostat from room
temperature down to 20~K, and diffraction patterns were then collected every
two minutes at a warming rate of 1~K/min from 20~K to 320~K. A wavelength of
$\sim 2.41$~{\AA} was used to highlight the magnetic diffraction and was
calibrated using a Silicon sample.

In addition, high--resolution NPD data were collected at diffractometer
Super--D2B of ILL for samples with $x_{\text{Ca}}=0,0.05$ and 0.10. A
wavelength of $\sim 1.594$~{\AA}  was used to collect patterns at selected
temperatures for approximately 3~h. It is worth noting that the volume of
sample available was not as much as the ideal for this kind of experiment, so
we looked for a compromise between the collection time, the available
beamtime, and our capabilities for preparing all the samples in a single
batch. The NPD patterns were processed with the full--pattern analysis
Rietveld method, using the program
\begin{scriptsize} FULLPROF
\end{scriptsize}~\cite{fullprofB} for refining the crystal and
magnetic structures.

\begin{figure}[t]
\centering \vspace{3mm}
\includegraphics[angle=-90,width=\linewidth]{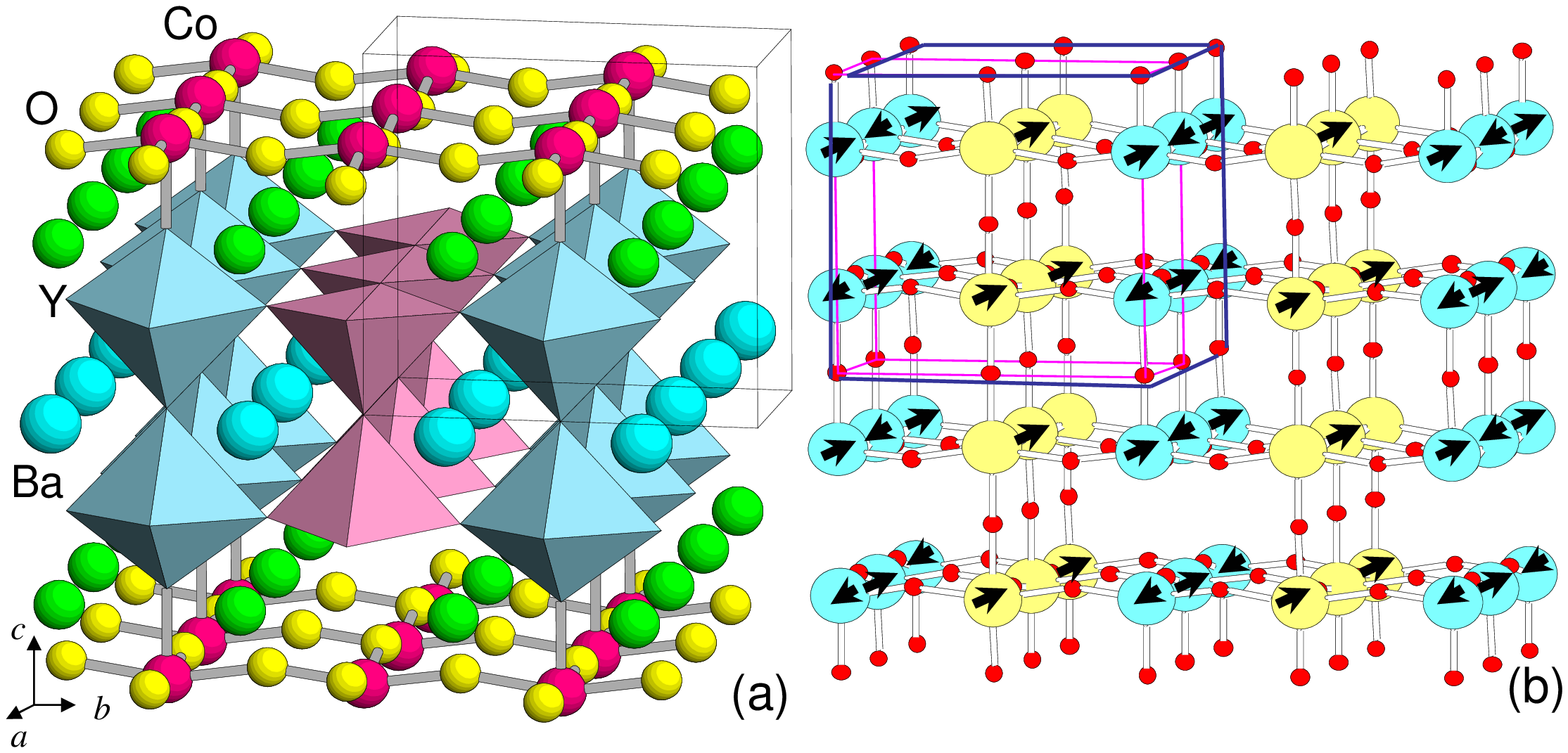}
\includegraphics[angle=-90,width=\linewidth]{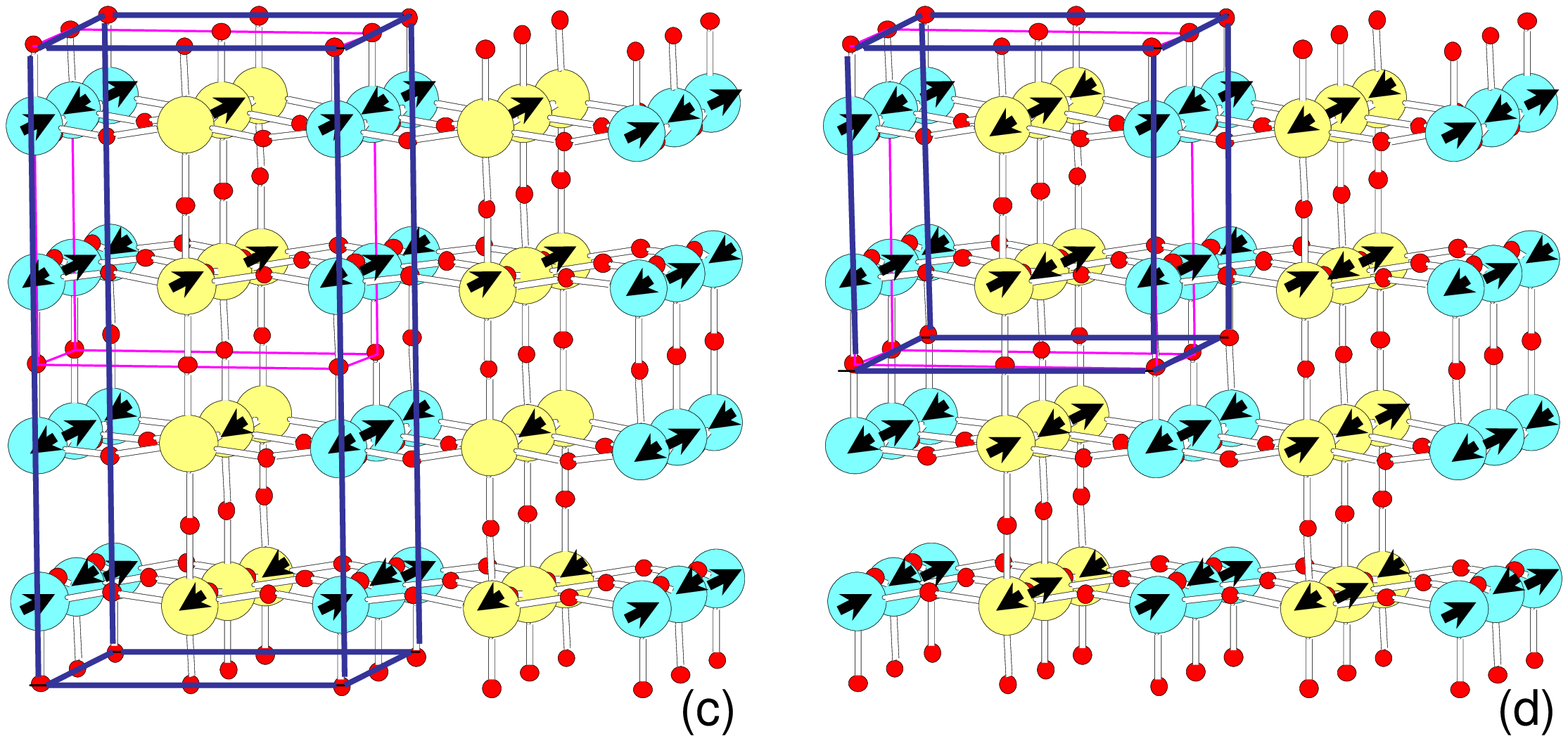}
\caption{ (Color online) (a) The structure of YBaCo$_{2}$O$_{5.5}$. Half Co
atoms are in square--pyramidal coordination and the other half in octahedral
coordination. (b) Magnetic model adopted for the SSO ferrimagnetic phase after
Ref.~\onlinecite{05Kha}. The nuclear ``122" unit cell and the magnetic ``222"
unit cell are indicated, and light atoms correspond to octahedral
coordination. In this schematic representation, only Co and O atoms are shown
for clarity. (c) Magnetic model adopted for the AFM--\emph{O}2 phase after
Ref.~\onlinecite{05Kha}. (d) Magnetic model adopted AFM--\emph{O}1 phase. }
\label{f:estructura}
\end{figure}

\section{Results \label{s:results}}

\subsection{Description of structures and refinement strategy} \label{ss:structures}

The room temperature structures of the parent compound YBaCo$_{2}$O$_{5+
\delta}$ were first reported by Akahoshi and Ueda,~\cite{99Aka} who showed
that for $\delta=0.5$ there may form two competing structures. One of them is
orthorhombic, and corresponds to the space group $Pmmm$ having the ``122"
superstructure characteristic of similar cobaltites with
$\delta=0.5$.~\cite{00Mor,01Kus,01Res,03Kha} A schematic representation of
this phase is shown in Fig.~\ref{f:estructura}. The vacancies order consists
of alternating [CoO$_{6}$] octahedra chains along the $c-$axis and
corner--sharing [CoO$_{5}$] pyramids along the $b-$axis, resulting in
alternating octahedral and pyramidal layers in the $a-c$ plane. This produces
a doubling of the cell along the $b-$axis, with a unit cell
$a_{P}$x$2a_{P}$x$2a_{P}$. The second structure that may stabilize in this
system for $\delta=0.5$ (and other values as well) has a tetragonal symmetry
and no doubling of the $b-$axis, \textit{i.e.}, no ordering between pyramids
and octahedra. In this case, the space group is $P4/mmm$ and the unit cell
$a_{P}$x$a_{P}$x$2a_{P}$. Recently, Frontera \textit{et al.}~\cite{05Fro} have
shown that, although the order of vacancies may not always be perfectly
achieved, the order between the \emph{R}--cation layer and the Ba layer is
well established and there is no mixing between them. Neither are there
significant oxygen vacancies in the [BaO] layers. The ``122" structure admits
a certain degree of disorder, consisting of misplaced pyramids or octahedra,
but keeping the long--range ``122" order.

In the present refinements, the Ca cations were randomly introduced in the
structure at the crystallographic site occupied by Ba, in appropriate
proportions. We have found no evidence of Ca segregation nor the formation of
additional phases, so we believe that Ca has been successfully incorporated
into the cobaltites structure. The strategy for the Rietveld refinement was as
follows. First, the high resolution data from D2B were refined to obtain an
accurate nuclear structure for each sample. The raw data coming from the
Super--D2B detector were processed using the LAMP software \cite{LAMP} to
obtain two sets of data: one of them having a better angle resolution at the
expense of losing some neutron counts, the other one having all neutron counts
collapsed into a single diffractogram. The first set was used to determine the
lattice parameters, while the second set was used to refine the atomic
positions and temperature factors, and both sets were iteratively refined
until convergence to the structure. The magnetic structures were also included
in the refinements. The models we have used will be discussed in the following
sections. At a second step, the structural data obtained were used to refine
sequentially the neutron thermodiffractograms obtained at D20. Temperature
scans where divided into different ranges according to the structural and
magnetic order, and for each range the atomic positions and occupations
obtained at D2B were kept fixed, while lattice parameters, temperature factors
and magnetic moments were allowed to vary.

The objective of this work is to focus on the role of Ca addition to the
parent compound YBaCo$_{2}$O$_{5.5}$. Our results will show that there is a
clear logical sequence between the three samples studied, corresponding to
$x_{\rm{Ca}}=0$, $x_{\rm{Ca}}=0.05$ and $x_{\rm{Ca}}=0.10$. Surprisingly, the
sample with greater Ca content, $x_{\rm{Ca}}=0.10$, turned out to be the
simplest one, and as Ca is removed the complexity increases, resulting in a
quite complicated temperature evolution of the parent compound. This fact
probably explains why this compound has not yet been fully reported in such
detail as other \emph{R} cobaltites, except for the structural study by
Akahoshi and Ueda,~\cite{01Aka} and a recent neutron diffraction study by
Khalyavin \textit{et al}.~\cite{07Kha} focusing on the ferrimagnetic to
paramagnetic transition. For the above reasons, we will present our results
following the decreasing Ca sequence.

\subsection{Sample with $x_{\rm{Ca}}=0.10$} \label{ss:10Ca}

\begin{figure}[b]
\centering
\includegraphics[angle=-90,width=\linewidth]{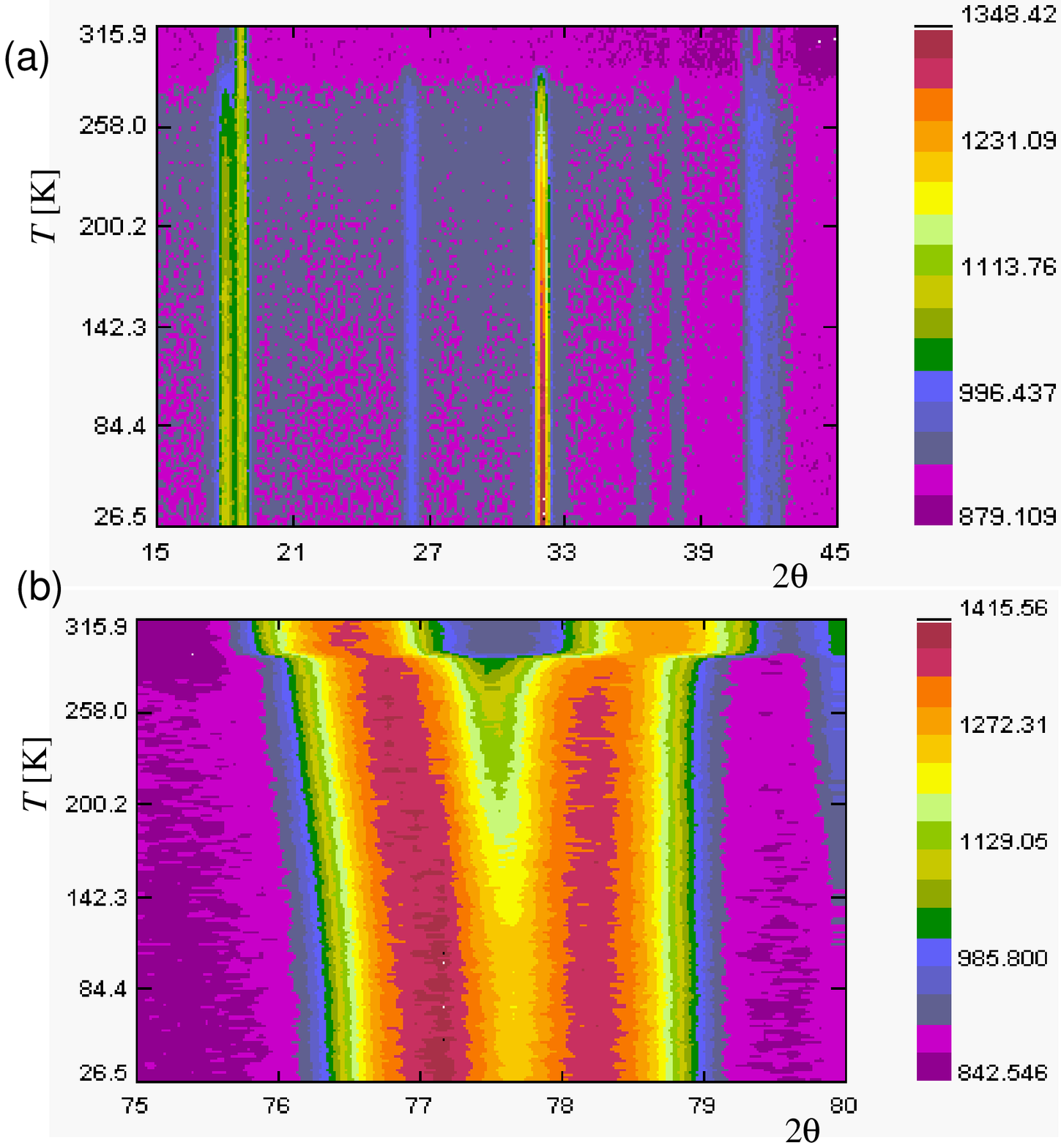}
\caption{(Color online) Projection of two selected sections of the
thermodiffractograms corresponding to sample with $x_{\rm{Ca}}=0.10$. (a)
$16^{\circ}< 2\theta < 44^{\circ}$ and (b) $75^{\circ}< 2\theta < 80^{\circ}$
showing the (2~0~0) and (0~4~0) Bragg reflections of the ``122" structure.
Data were collected at D20 with $\lambda \sim 2.41$ {\AA} between 20~K and
320~K.} \label{f:D20-X}
\end{figure}

In Fig.~\ref{f:D20-X} we present two different sections of the projected
thermodiffractograms for the sample  $x_{\rm{Ca}}=0.10$. Fig.~\ref{f:D20-X}(a)
corresponds to the low--angle range, in which most reflections are of magnetic
nature, and disappear simultaneously at $ \sim 295$~K on warming from 20~K to
300~K. In  Fig.~\ref{f:D20-X}(b) we focus on the $2 \theta$ range where the
Bragg reflections (2~0~0) and (0~4~0) clearly show a distortion occurring at
room temperature.

The high resolution data were refined using a nuclear phase with the ``122"
structure, as described in section~\ref{ss:structures}. We do not discard the
possibility of the actual structure being ``222", with a doubling of the
$a-$axis and four different crystallographic sites for the Co
atoms,~\cite{05Kha,07Kha} in line with the magnetic model adopted. However, as
we are interested in the temperature evolution of rather low resolution data
from D20, and given the complexity of the other samples, we have decided to
refine the whole series with the averaged ``122" structure. Moreover,
following Frontera \textit{et al.}~\cite{06Fro} we have fixed the \emph{z}
coordinate of the octahedral Co site to 1/4, in order to obtain
centrosymmetric octahedra. With these assumptions, we reduced the number of
free parameters which is critical when dealing with multiphasic systems. We
have nevertheless allowed for disorder between pyramids and octahedra, by
refining the occupation of apical oxygen sites (site \emph{1g} mostly
occupied, and site \emph{1c} mostly unoccupied in the \emph{Pmmm} space
group). We remark that even for different degrees of vacancies disorder, the
total occupation of sites always summed up to the same oxygen content in all
samples below 350~K. In Table~\ref{t:X} we present the details of the refined
structure for the sample with $x_{\rm{Ca}}=0.10$ from D2B data collected at
70~K, 230~K and 348~K.

\begin{table}[bt]
\caption{Structural parameters refined from the high resolution D2B data for
the compound YBa$_{0.90}$Ca$_{0.10}$Co$_2$O$_{5.5}$ at $T=70$~K, 230~K and
350~K. Atomic fractional coordinates correspond to space group $Pmmm$ in the
following Wyckoff positions: Y (\emph{2p})=($\frac{1}{2},y,\frac{1}{2}$); Ba,
Ca (\emph{2o})=($\frac{1}{2}, y, 0)$; CoOct
(\emph{2r})=($0,\frac{1}{2},\frac{1}{4}$); CoPyr (\emph{2q})= ($0,0,z$); O1
(\emph{1a})=($0,0,0$); O2 (\emph{1e})=($0,\frac{1}{2},0$); O3
(\emph{1g})=($0,\frac{1}{2},\frac{1}{2}$); O3'
(\emph{1c})=($0,0,\frac{1}{2}$); O4 (\emph{2s})=($\frac{1}{2},0,z$); O5
(\emph{2t})=($\frac{1}{2},\frac{1}{2},z$); O6 (\emph{4u})=($0,y,z$).}
\label{t:X}
\begin{ruledtabular}
\begin{tabular}{lcccc}
 & & $T=70$~K & $T=230$~K & $T=350$~K \\ \hline
Y & $y$ & 0.2714(5)  & 0.2717(5)   & 0.2684(5)  \\
Ba, Ca & $y$ & 0.253(1)  & 0.254(1)  & 0.248(1)  \\
CoPyr & $z$ & 0.260(1)  & 0.260(1)  & 0.262(1)  \\
O4 & $z$ & 0.3112(7)  & 0.3108(7)  & 0.3107(7)  \\
O5 & $z$ & 0.276(1)  & 0.276(1)  & 0.271(1)  \\
O6 & $y$ & 0.2481(8)  &  0.2479(8) & 0.2416(8)  \\
O6 & $z$ & 0.2954(6)  & 0.2947(5)  & 0.2980(5)  \\
O3 & $Occ$ &  0.94(2) & 0.93(2)  & 0.89(2)  \\
O3' & $Occ$ & 0.0  & 0.0  & 0.02(2)  \\
$a$ ({\AA}) & & 3.8423(1)  & 3.8438(1) & 3.8254(1) \\
$b$ ({\AA}) & & 7.7947(2)  & 7.8118(2) & 7.8503(2) \\
$c$ ({\AA}) & & 7.4835(2)  & 7.4965(2) & 7.5217(2) \\
$R_{B}$ &  & 7.4  & 7.4  & 7.1  \\
$R_{mag}$ &  & 13.6 &  14.7 & \\
$\chi^2$ & & 7.4 & 5.8 & 5.2  \\
\end{tabular}
\end{ruledtabular}
\end{table}

\begin{figure}[h]
\centering \vspace{3mm}
\includegraphics[angle=-90,width=.9\linewidth]{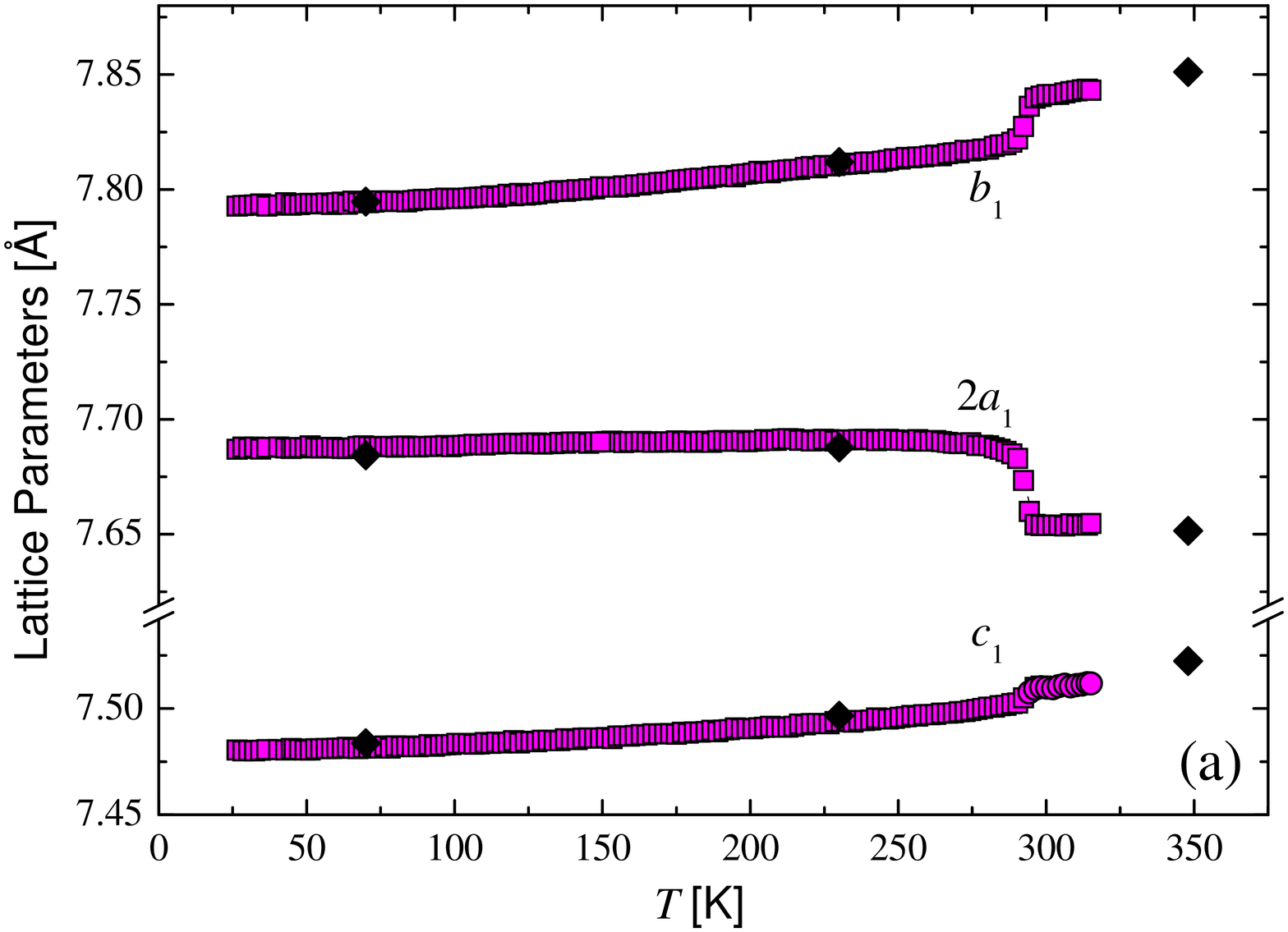}
\includegraphics[angle=-90,width=.9\linewidth]{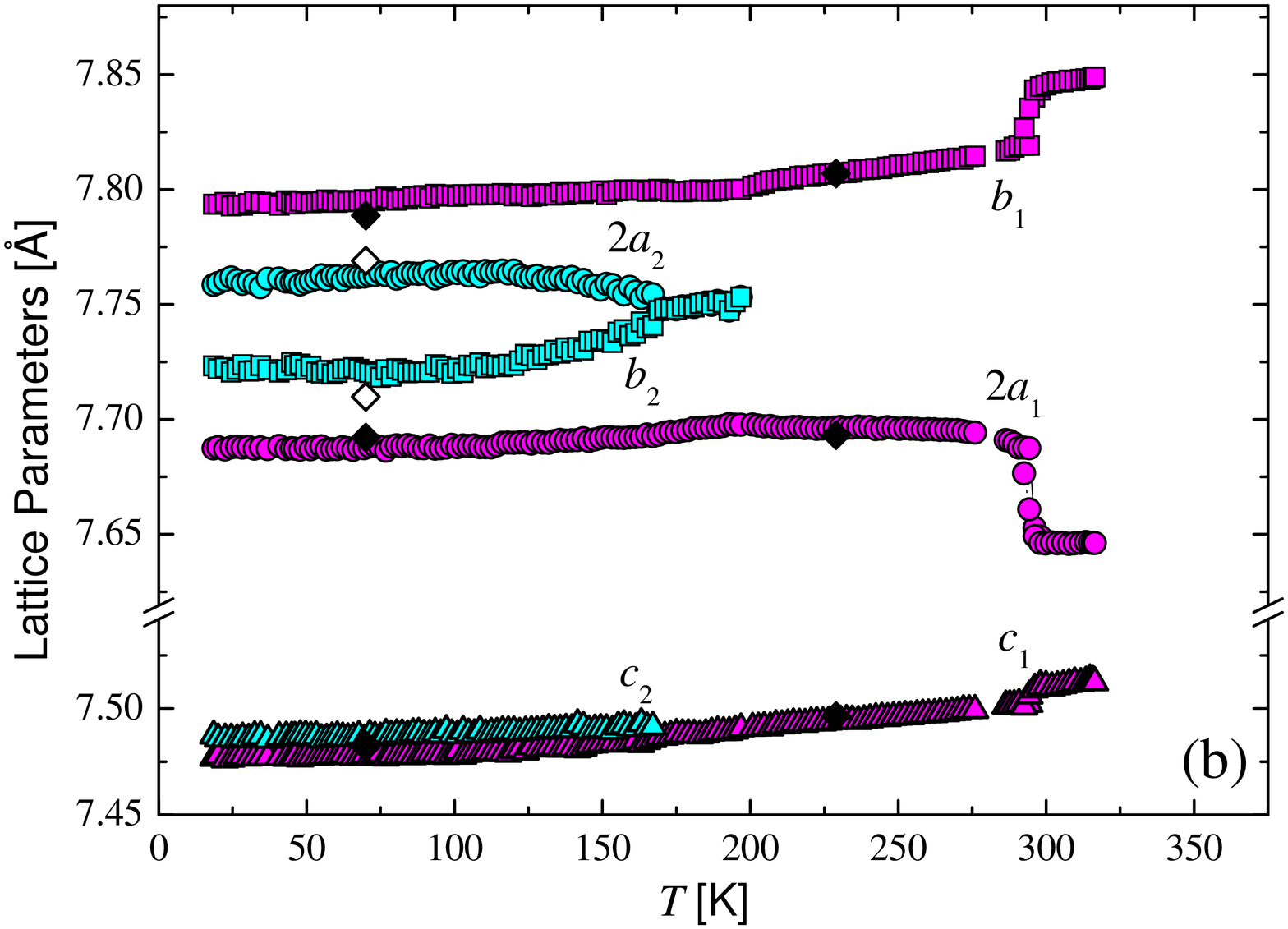}
\includegraphics[angle=-90,width=.9\linewidth]{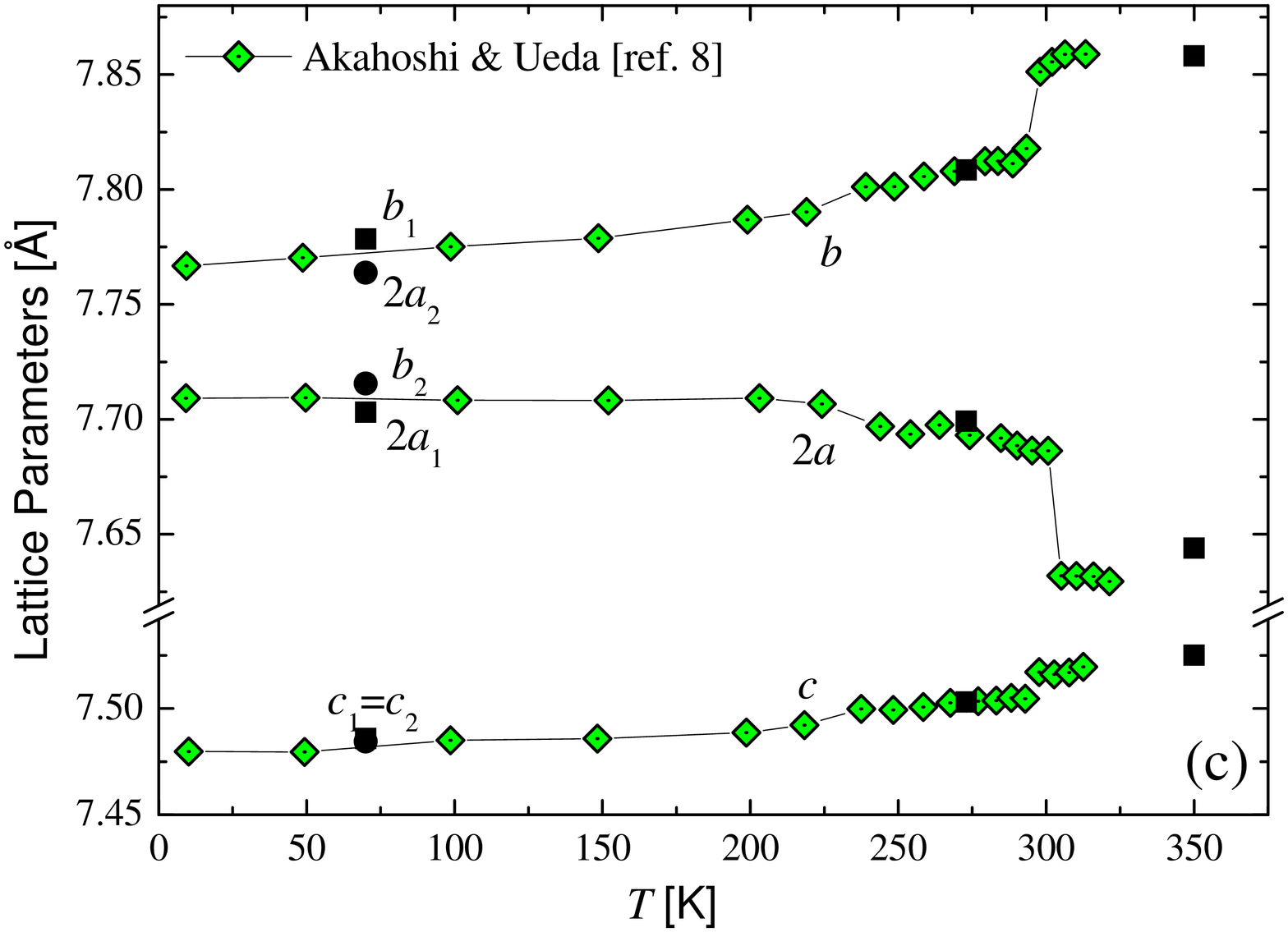}
\caption{(Color online) Thermal evolution of the lattice parameters $2a$, $b$
and $c$ for the \emph{O}1 (dark symbols) and \emph{O}2 (light symbols) phases
in samples with $x_{\text{Ca}}=0.10$ (a), $x_{\text{Ca}}=0.05$ (b) and
$x_{\text{Ca}}=0$ (c), determined from data collected at D20 and D2B. In (c)
the data from D2B (solid symbols) are compared with data reported by Akahoshi
and Ueda,~\cite{01Aka} plotted using diamonds.} \label{f:LPs}
\end{figure}

\begin{figure}[h]
\centering \vspace{3mm}
\includegraphics[angle=-90,width=.9\linewidth]{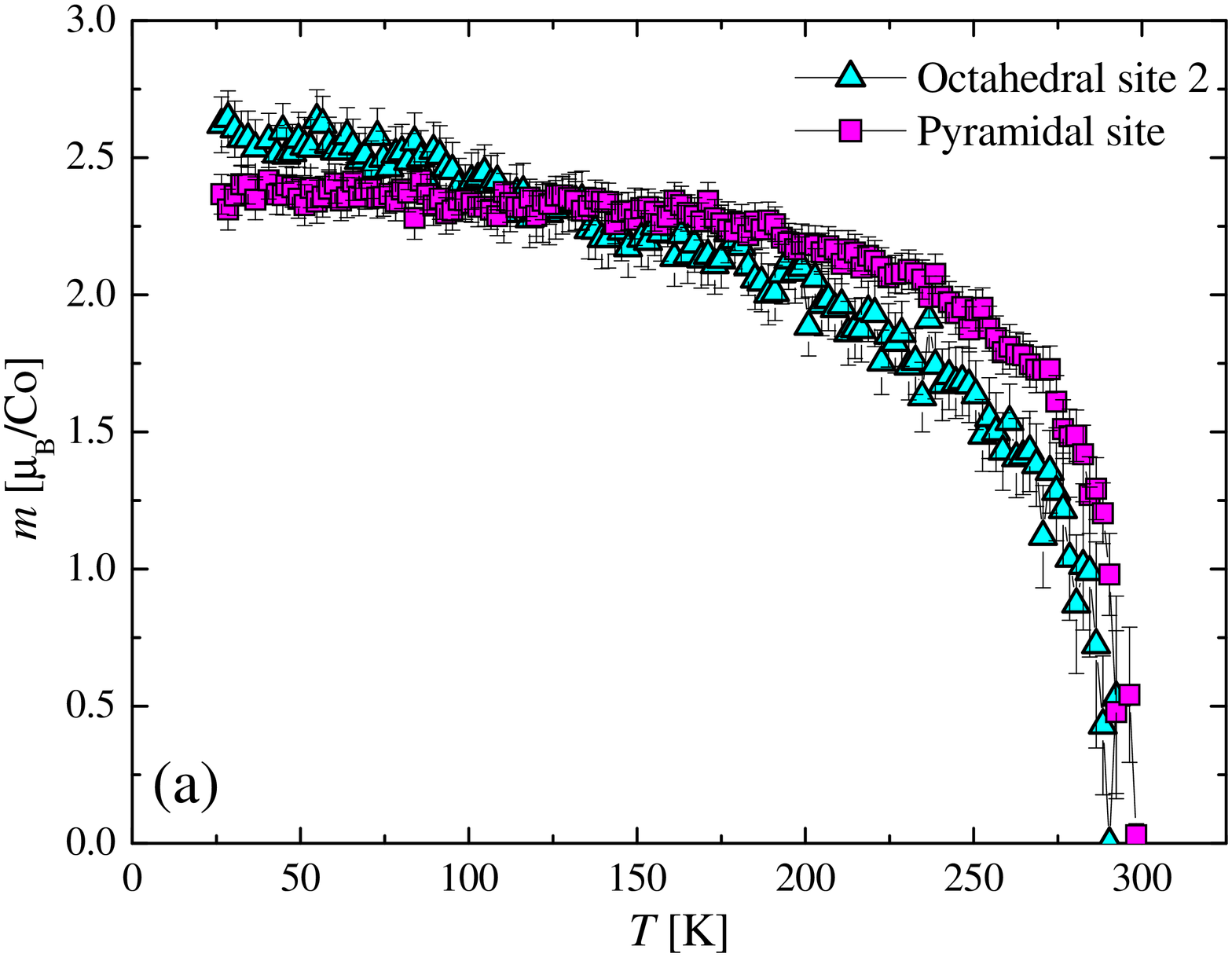}
\includegraphics[angle=-90,width=.9\linewidth]{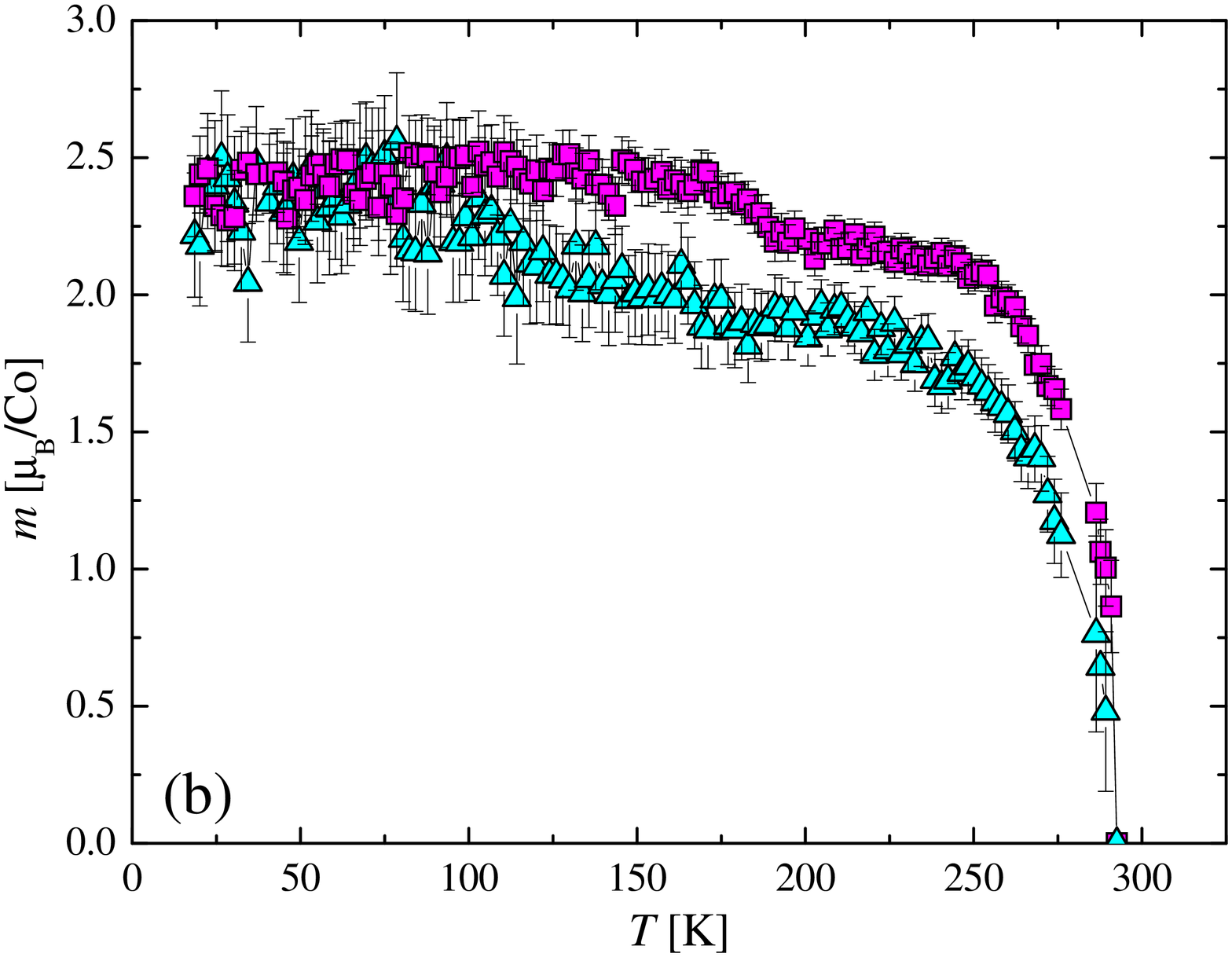}
\caption{(Color online) Temperature dependence of the magnetic moments in
pyramids (square symbols) and half the octahedra (triangular symbols) for the
samples $x_{\text{Ca}}=0.10$ (a) and $x_{\text{Ca}}=0.05$ (b), obtained from
the Rietveld refinements of neutron data collected at D20. Moments were
refined along [100]. } \label{f:mu-YBCoXIyX}
\end{figure}

The model we have adopted for refining the magnetic phase is the SSO
ferrimagnetic model proposed by Khalyavin,~\cite{05Kha,07Kha} which leads to
the best agreement with our NPD data, macroscopic magnetization data and high
temperature susceptibility data.~\cite{law3m,07Aur} It is schematized in
Fig.~\ref{f:estructura}(b). Although the model involves Co atoms at half the
octahedral sites being in low--spin state, and therefore having a magnetic
moment equal to zero, after a first step in the refinement we allowed this
site to adopt a non--zero magnetic moment, which resulted in a small value
compatible with the fact that the apical oxygen site \emph{1g} is not
completely occupied, and therefore some octahedra are, in fact, misplaced
pyramids.~\cite{06Fro} In Fig.~\ref{f:LPs}(a) we present the evolution with
temperature of the lattice parameters refined in the $Pmmm$ ``122" structure.
The sequential D20 results are presented together with the results from D2B at
the studied temperatures. The characteristic structural distortion occurring
at $T_{\rm{MI}}\sim 295$~K is clearly observed. We have already reported that
in this system, the MI transition occurs almost simultaneously with the
paramagnetic--ferrimagnetic transition,~\cite{law3m} which seems to be only a
coincidence. Therefore, no further anomaly is observed in the lattice
parameters down to 20~K, as there is in this sample no further magnetic
transition. In Fig.~\ref{f:mu-YBCoXIyX}(a) we present the magnetic moment of
Co atoms in each crystallographic environment. We have not included in the
figure the small magnetic moment of misplaced pyramids, which remains always
less than 0.4 $\mu_{\rm{B}}$.

\begin{figure}[h]
\centering \vspace{3mm}
\includegraphics[angle=-90,width=\linewidth]{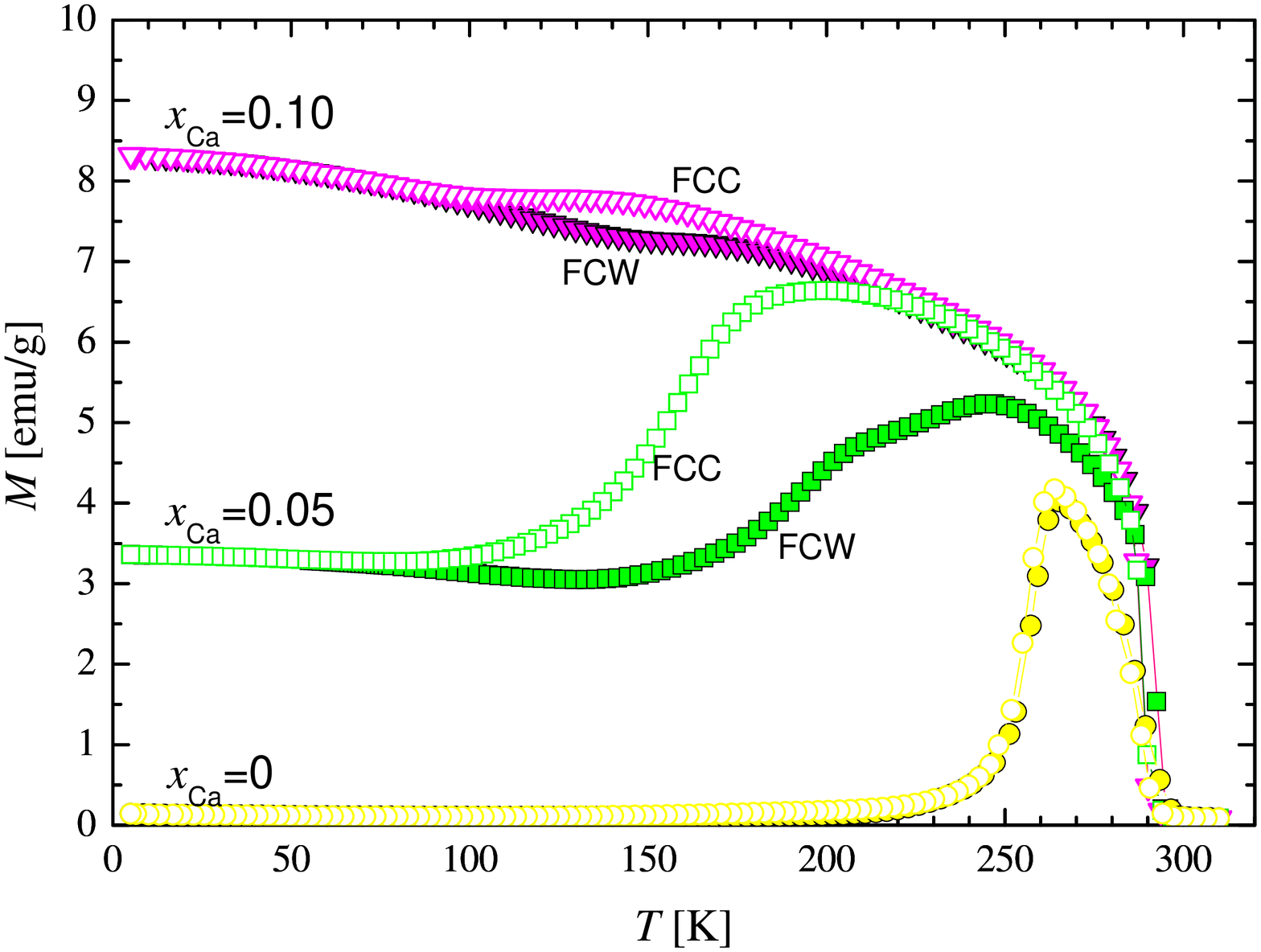}
\caption{(Color online) Low--field magnetization as a function of temperature
for samples with $x_{\text{Ca}}=0$, $x_{\text{Ca}}=0.05$ and
$x_{\text{Ca}}=0.10$. Empty symbols represent the magnetization measured on
cooling under a magnetic field of 5~kOe (FCC), and solid symbols represent the
magnetization subsequently measured on warming (FCW).} \label{f:MdeT}
\end{figure}

\subsection{Sample with $x_{\rm{Ca}}=0.05$} \label{ss:5Ca}

Our preliminary X--ray diffraction pattern of the as--synthesized sample
$x_{\text{Ca}}=0.05$ at room temperature resulted identical to
$x_{\text{Ca}}=0.10$ with just a slight difference in the lattice parameters,
indicating that the room temperature structure is the same in both samples.
However, the macroscopic magnetization data in Fig.~\ref{f:MdeT} show that at
low temperature they behave differently. Below 200~K, on cooling, the
magnetization of the $x_{\text{Ca}}=0.05$ sample starts dropping but the
sample retains a net magnetization down to 5~K, in contrast with the parent
compound $x_{\text{Ca}}=0$ which shows an AFM behavior. Moreover, the big
hysteresis between the data collected on cooling and warming in the
$x_{\text{Ca}}=0.05$ sample suggests that there is a competition between two
states. The NPD experiments reveal the nature of these two states.

\begin{figure*}[t]
\centering \vspace{3mm}
\includegraphics[angle=-90,width=\linewidth]{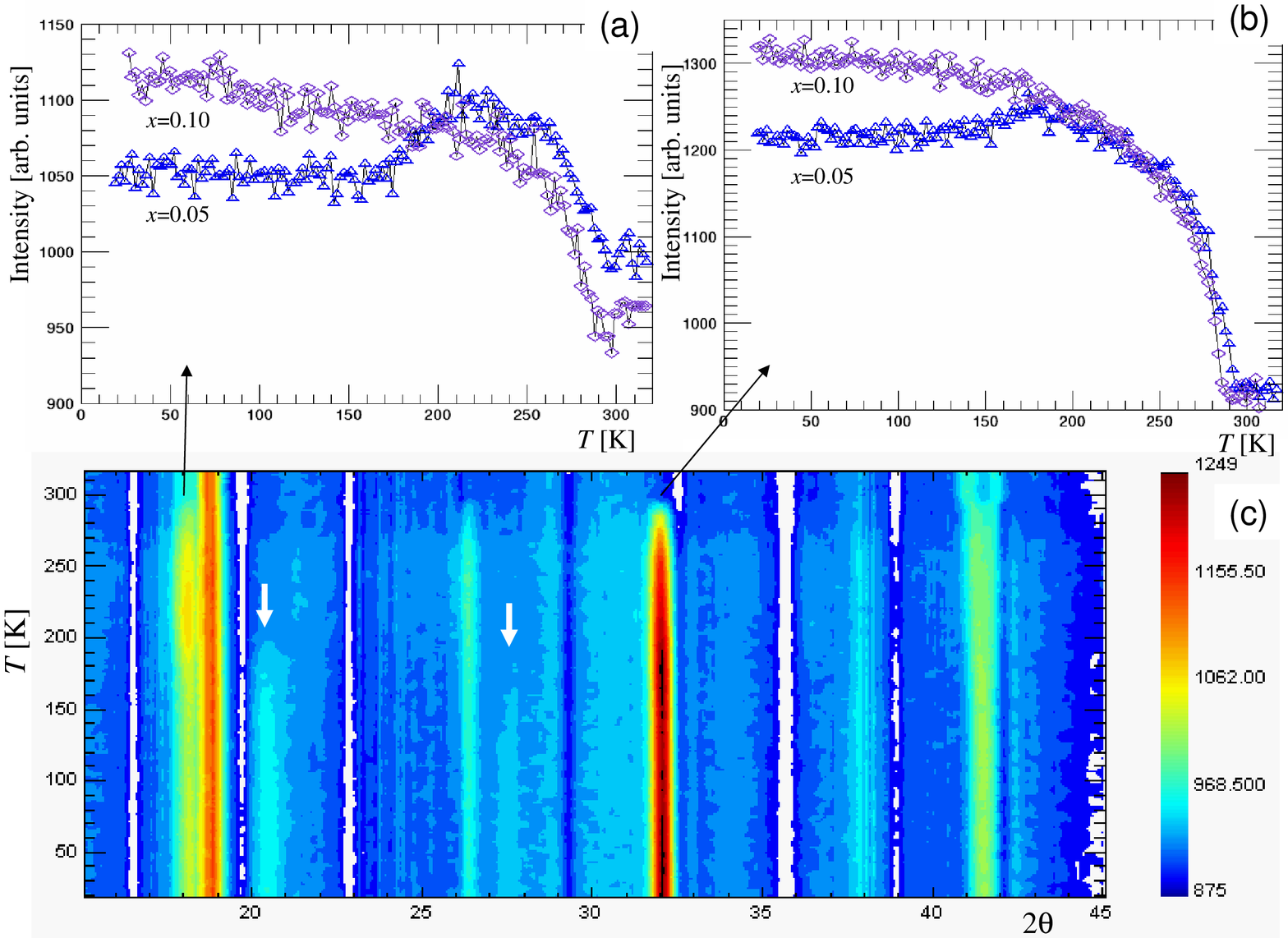}
\caption{(Color online) Thermal evolution of the intensity of Bragg
reflections (0 1 0)(a) and (1/2 1 1)(b) indexed in terms of the ``122" phase
for samples with $x_{\text{Ca}}=0.05$ (triangles) and $x_{\text{Ca}}=0.10$
(diamonds) from data collected at D20. (c) Two--dimensional projection of the
neutron thermodiffractograms collected at D20 for sample $x_{\text{Ca}}=0.05$
in the range $16^{\circ}<2\theta<44^{\circ}$. The small arrows indicate the
onset of magnetic reflections from the AFM ``224" phase. The long arrows
between graphs indicate the position in the thermodiffractogram of the two
reflections whose intensity is plotted in (a) and (b).}
\label{f:picos-magneticos}
\end{figure*}

\begin{figure}[t]
\centering \vspace{3mm}
\includegraphics[angle=-90,width=\linewidth]{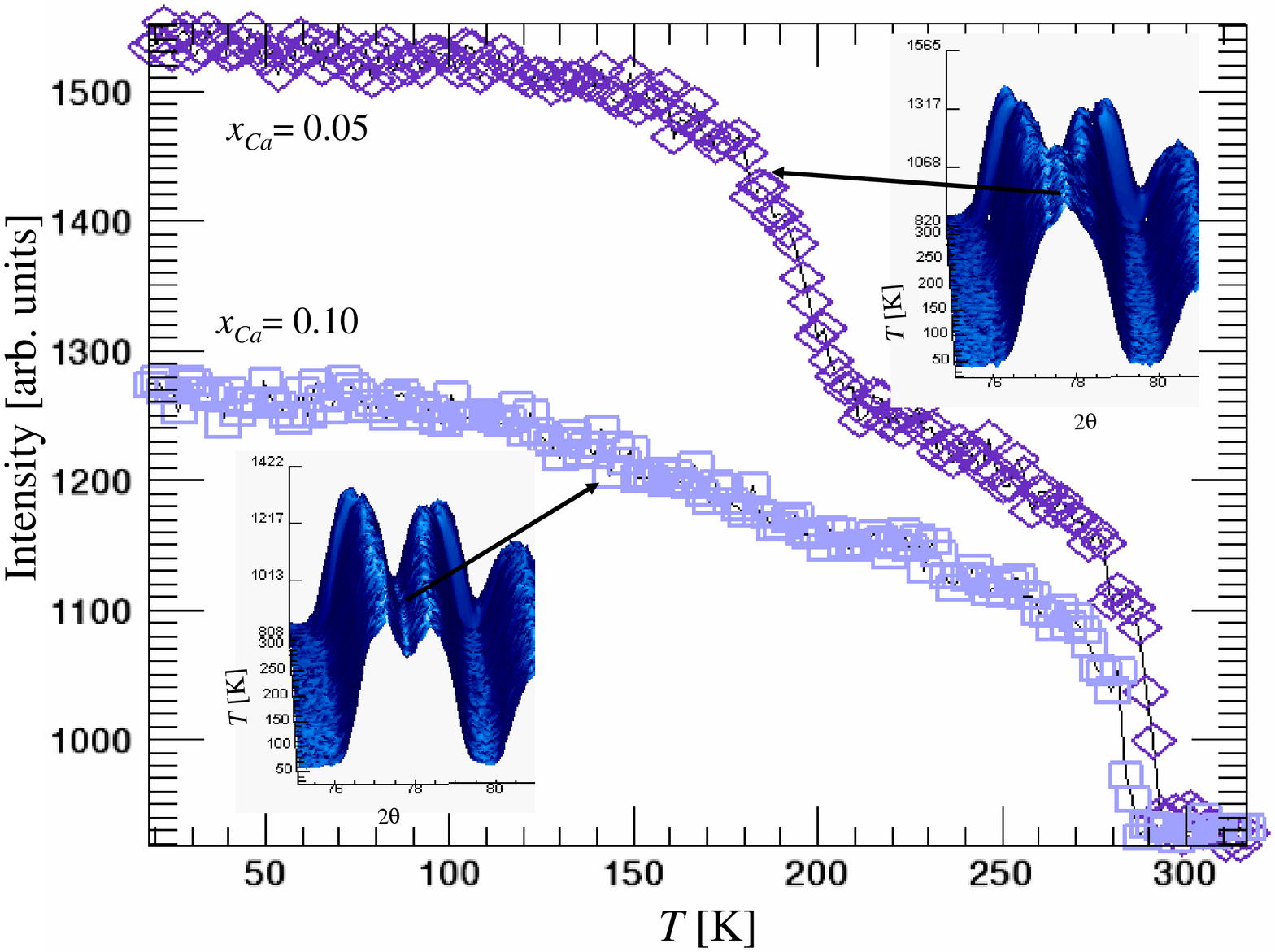}
\caption{(Color online) Thermal evolution of the intensity in the
thermodiffractograms at $2\theta \sim 77.6^{\circ}$, in the position between
the (2~0~0) and (0~4~0) Bragg reflections of the ``122" phase for the sample
with $x_{\text{Ca}}=0.05$ (diamonds) and $x_{\text{Ca}}=0.10$ (squares). The
insets show the three--dimensional thermodiffractograms for the relevant
$2\theta$ range. } \label{f:pico-tetragonal}
\end{figure}

In Fig.~\ref{f:picos-magneticos} (c) we present a projection of the
thermodiffractograms collected at D20 for sample $x_{\text{Ca}}=0.05$. In (a)
and (b) we show the temperature evolution of the intensity of the magnetic
reflections (0 1 0) and (1/2 1 1) respectively, indexed in terms of the ``122"
nuclear phase. These reflections correspond to the ferrimagnetic order
discussed in the previous subsection. Figure~\ref{f:picos-magneticos} (a) and
(b) compare the intensities of these reflections in samples
$x_{\text{Ca}}=0.05$ and $x_{\text{Ca}}=0.10$. We can observe that they both
behave almost identically from 200~K to 320~K, but below 200~K the
ferrimagnetic reflections are lower in the $x_{\text{Ca}}=0.05$ sample.
Moreover, at 200~K additional features are evidenced in the
thermodiffractograms. The small arrows in Fig.~\ref{f:picos-magneticos} (c)
mark two reflections which appear only below 200~K. They are indicative of the
presence of a second ---magnetically ordered--- phase which may be indexed
with a further doubling of the $c$ parameter as reported by various authors in
the AFM region of layered cobaltites.~\cite{02Fau,05Kha,05Pla,06Fro,03Sod} But
here we also observe simultaneous changes in the nuclear structure and not
only a magnetic phase separation, or gradual magnetic transition to a
different magnetic arrangement. ~\cite{02Fau,05Pla,06Fro,06Lob} This is
illustrated, for instance, by a peak arising at 200~K in $2\theta \sim
77.6^{\circ}$ lying in between the (2~0~0) and (0~4~0) reflections of the
``122" nuclear phase. In Fig.~\ref{f:pico-tetragonal} we show the evolution
with temperature of the collected intensity at $2\theta \sim 77.6^{\circ}$,
for this sample as well as for the sample with $x_{\text{Ca}}=0.10$ for
comparison. When warming, there is a sudden drop at 200~K, to a value
corresponding to the overlap of the neighboring (2~0~0) and (0~4~0)
reflections, and a further drop to background values above the MI transition,
when these reflections are suddenly shifted apart by the distortion. We first
evaluated the possibility of this behavior being a result of a distortion of
the ``122" nuclear phase at 200~K. Such hypothesis gave no satisfactory
results when trying to refine simultaneously the D2B and D20 data at 70~K.
Additional attempts to model the $70^{\circ} < 2 \theta < 80^{\circ}$ region
using Gaussian peaks of just one orthorhombic phase, even allowing for
unrealistically wide peaks, could not reproduce the triple--peak shape
observed in the D20 spectra (inset in Fig.~\ref{f:pico-tetragonal}).

We also considered other structural models to account for the low temperature
data coming from D2B and D20. Neither the $Pcca$ nor $Pmma$ space groups
proposed by Plakhty \emph{et al.}~\cite{05Pla} and Khalyavin,~\cite{07Kha}
together with the respective magnetic models proposed in their work, gave
satisfactory results for the simultaneous refinement of all our data.

 Another possible explanation is the presence of a second
structural phase. This is suggested by the poor results obtained refining the
D2B data using the same (single-phase) model as for the $x_{\text{Ca}}=0.10$
sample at 70~K ($Pmmm$ space group), as well as with other single-phase models
($Pcca$, $Pmma$). Based on the D20 data, we considered a possible second phase
with a strong tetragonal distortion, which would not be unreasonable
considering that a two-phase mixture of orthorhombic and tetragonal phases for
YBaCo$_2$O$_{5.5}$ had already been reported by Akahoshi and
Ueda,~\cite{99Aka} both phases occurring competitively. It is also worth
noting that in the same batch as the present samples, we have also synthesized
a series of samples where Barium is replaced by Strontium, a substitution
which clearly favors a tetragonal phase in which the position of the (2 0 0)
Bragg reflection is almost exactly coincident with this new peak in the
$x_{\text{Ca}}=0.05$ sample.~\cite{inprogress} Consequently, we tried to
refine both the D2B and D20 sets of data using a mixture of the ``122" phase
and a tetragonal ($P4/mmm$) phase. This gave no satisfactory results either,
and moreover, the proposed tetragonal symmetry could not account for the
observed magnetic supercell.

\begin{figure}[tb]
\centering \vspace{3mm}
\includegraphics[angle=-90,width=\linewidth]{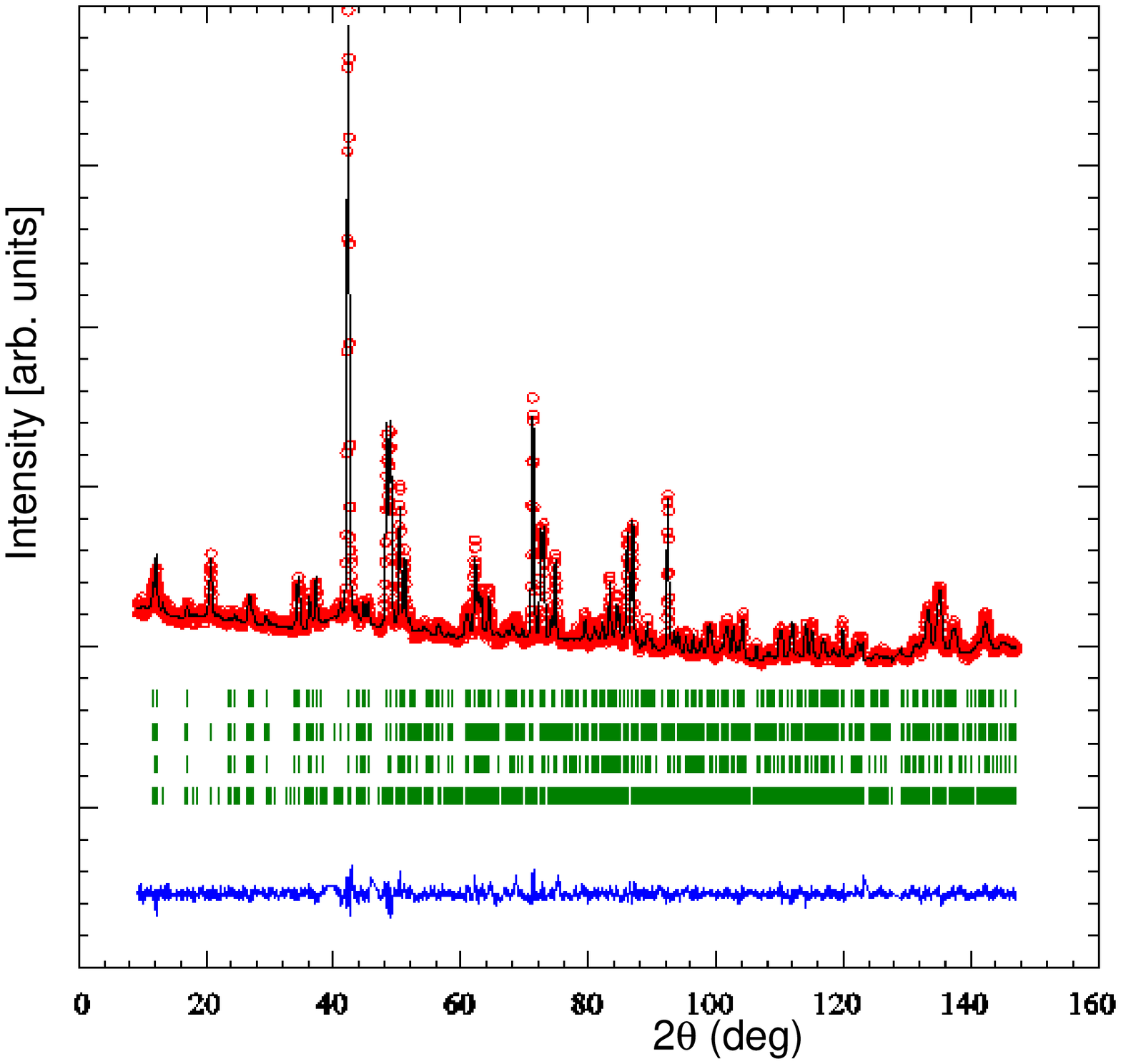}
\caption{(Color online) Rietveld refinement for the sample with
$x_{\text{Ca}}=0.05$ from data collected at D2B at $T=70$~K. Vertical bars at
the bottom indicate Bragg reflections from the phases included in the
refinement: the nuclear phases \emph{O}1, \emph{O}2 and the magnetic phases
SSO ``222" and AFM ``224". } \label{f:prf-IX-70K}
\end{figure}

We finally proceeded to adopt for the second phase an orthorhombic ($Pmmm$)
cell with the constraint $b=2a$, and at a further step we allowed the $b$ and
$a$ lattice parameters to vary independently. The diffractogram at 70~K was
therefore refined with one orthorhombic phase with ferrimagnetic order
(\emph{O}1 phase) plus a second orthorhombic phase (\emph{O}2) with AFM order
which is consistent with a ``224" supercell. This finally gave much more
satisfactory results for the refinement. The magnetic peaks arising below
200~K, although weak, could be accounted for using the AFM model proposed by
Khalyavin,~\cite{05Kha} schematized in Fig.~\ref{f:estructura}(c), with the
constraint of a single magnitude for the magnetic moment of all Co atoms in
pyramidal positions, and once again (as in the ferrimagnetic SSO model), two
possible spins for octahedral Co. At 230~K, the \emph{O}1 phase with
ferrimagnetic order was enough to refine the D2B diffractogram. In
Table~\ref{t:IX} we present the details of the refined structures for the
sample with $x_{\rm{Ca}}=0.05$ from D2B data collected at 70~K and 230~K.
Figure~\ref{f:prf-IX-70K} shows the Rietveld refinement (solid line) of the
high resolution data (symbols) collected at 70~K. The four sets of Bragg
reflections indicated at the bottom by vertical bars correspond to each of the
above mentioned phases. The difference pattern between observed and calculated
data is also shown.

The presented scenario for the evolution with temperature of the
$x_{\text{Ca}}=0.05$ sample yields consistent results among the D2B and D20
data, as well as it accounts for other experimental facts. When cooling from
room temperature, the intensity of the ferrimagnetic reflections starts
dropping because there is a second phase developing in the sample, so that the
volume fraction of the ferrimagnetic phase is reduced. In addition, the
observed hysteresis among the FCC and FCW magnetization curves in
Fig.~\ref{f:MdeT} are also indicative of a possible phase separation, as well
as the hysteresis in the resistivity curves presented previously.~\cite{07Aur}
The presence of a second nuclear phase has become more evident when performing
thermodiffractograms with $\lambda=2.52 {\AA}$. This scenario would be very
difficult to infer just from D2B data collected at lower wavelengths and at
isolated temperatures, due to peak overlap and to a lack of perspective of the
continuous thermal evolution of the sample.

\begin{table}[bt]
\caption{Structural parameters refined from the high resolution D2B data for
the compound YBa$_{0.95}$Ca$_{0.05}$Co$_2$O$_{5.5}$ at $T=70$~K and 230~K. Two
sets of atomic fractional coordinates are given for the \emph{O}1 and
\emph{O}2 phases, which correspond to space group $Pmmm$ in the following
Wyckoff positions: Y (\emph{2p})=($\frac{1}{2},y,\frac{1}{2}$); Ba, Ca
(\emph{2o})=($\frac{1}{2}, y, 0)$; CoOct
(\emph{2r})=($0,\frac{1}{2},\frac{1}{4}$); CoPyr (\emph{2q})= ($0,0,z$); O1
(\emph{1a})=($0,0,0$); O2 (\emph{1e})=($0,\frac{1}{2},0$); O3
(\emph{1g})=($0,\frac{1}{2},\frac{1}{2}$); O3'
(\emph{1c})=($0,0,\frac{1}{2}$); O4 (\emph{2s})=($\frac{1}{2},0,z$); O5
(\emph{2t})=($\frac{1}{2},\frac{1}{2},z$); O6 (\emph{4u})=($0,y,z$).}
\label{t:IX}
\begin{ruledtabular}
\begin{tabular}{lcccc}
 & & \multicolumn{2} {c} {$T=70$~K}& $T=230$~K \\
 &  &  \emph{O}1 & \emph{O}2 & \emph{O}1 \\ \hline
Y & $y$ & 0.2757(8)  & 0.264(2)   & 0.2731(5)  \\
Ba, Ca & $y$ & 0.254(1)  & 0.233(3)  & 0.254(1)  \\
CoPyr & $z$ & 0.262(2)  & 0.267(2)  & 0.261(1)  \\
O4 & $z$ & 0.317(1)  & 0.291(5)  & 0.3125(7)  \\
O5 & $z$ & 0.261(2)  & 0.287(5)  & 0.274(1)  \\
O6 & $y$ & 0.2460(9)  &  0.243(2) & 0.2473(7)  \\
O6 & $z$ & 0.2941(6)  & 0.301(1)  & 0.296(1)  \\
O3 & $Occ$ &  1.0 & 0.84(6)  & 0.92(2)  \\
O3' & $Occ$ & 0.0  & 0.0  & 0.00(2)  \\
$a$ ({\AA}) & & 3.8460(2) & 3.8845(3) & 3.8468(1)  \\
$b$ ({\AA}) & & 7.7887(4) & 7.7099(6) & 7.8075(2)  \\
$c$ ({\AA}) & & 7.4827(5) & 7.4819(7) & 7.4959(2)  \\
$f$ (\%) & & 64(4) & 36(3) & 100 \\
$R_{B}$ &  & 7.0  & 7.0  & 7.3  \\
$R_{mag}$ &  & 15 & 32  & 16  \\
$\chi^2$ & &  & & 4.48  \\
\end{tabular}
\end{ruledtabular}
\end{table}

The evolution with temperature of the lattice parameters refined in the
\emph{O}1 and \emph{O}2 phases is shown in Fig.~\ref{f:LPs}(b). The sequential
D20 results are presented together with the results from D2B at the studied
temperatures. For the \emph{O}1 phase, the distortion at $T_{\rm{MI}}\sim
295~K$ is again observed as in the sample with $x_{\text{Ca}}=0.10$. For the
\emph{O}2 phase, we observe a tetragonal distortion above 170~K: above that
temperature the $a_2$ and $b_2$ lattice parameters could only be refined using
the constraint $2a_2=b_2$. The volume per atom of both phases is practically
the same, although it is observed that the lattice parameter relation is
different ($2a_1 < b_1$, $2a_2 > b2$). It would be interesting to stabilize
the \emph{O}2 phase to get more detailed information of its structural
properties.~\cite{inprogress} In Fig.~\ref{f:mu-YBCoXIyX}(b) we present the
magnetic moment of Co atoms in the ferrimagnetic phase for each
crystallographic environment. We have not included in the figure the small
magnetic moment of misplaced pyramids, which remains always less than 0.5
$\mu_{\rm{B}}$. Unfortunately, the quality of our D20 data and the two-phase
scenario do not allow for a confident determination of the magnetic moments in
the \emph{O}2 AFM phase. These were constrained to be aligned along [100]
following the model described above and to adopt values similar to those in
the \emph{O}1 phase, in order to obtain reliable phase fractions which were
comparable to the values refined from D2B data.

\begin{figure}[tb]
\centering \vspace{3mm}
\includegraphics[angle=-90,width=\linewidth]{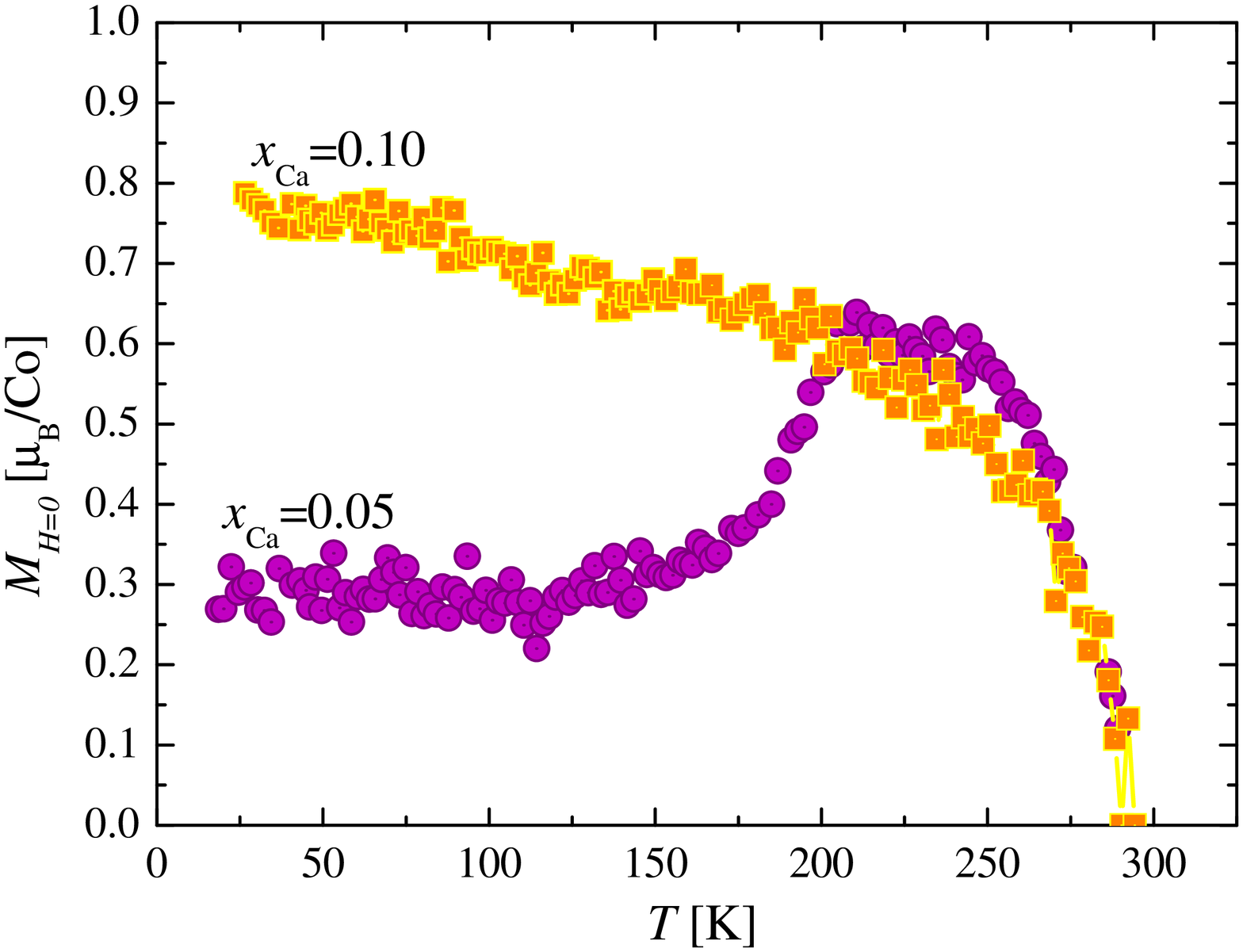}
\caption{(Color online) Spontaneous magnetization of the ferrimagnetic phase
in samples with $x_{\rm{Ca}}=0.05$ and $x_{\rm{Ca}}=0.10$, obtained from our
NPD refinements as $M=\mu_{\mathrm{Co}} \cdot f_{O1}$, where
$\mu_{\mathrm{Co}}$ represents the net magnetic moment per Co atom
($=\mu_{\mathrm{CoOct}}/4$) and $f_{O1}$ is the phase fraction of the
\emph{O}1 phase. } \label{f:magnet-neutrones}
\end{figure}

Figure~\ref{f:magnet-neutrones} shows the net spontaneous magnetization of
samples with $x_{\rm{Ca}}=0.05$ and $x_{\rm{Ca}}=0.10$, obtained as
$M=\mu_{\mathrm{Co}} \cdot f_{O1}$, where $\mu_{\mathrm{Co}}$ represents the
net magnetic moment per Co atom in the ferrimagnetic phase
($=\mu_{\mathrm{CoOct}}/4$) and $f_{O1}$ is the refined phase fraction of the
\emph{O}1 phase. These results can be compared with the macroscopic
determination of the magnetization of the samples as a function of temperature
(Fig.~\ref{f:MdeT}), always considering these were collected under an applied
field of 5~kOe. The overall similarity between the curves is in excellent
agreement.

\subsection{Sample with $x_{\rm{Ca}}=0$} \label{ss:PC}

We finally turn to the parent compound. This sample was only studied in the
high resolution instrument, so we cannot present continuous temperature scans
in the low--temperature range as in the other samples. We collected three
diffractograms at $T=70$~K, 273~K and 348~K. A similar study on this system
has been very recently reported,~\cite{07Kha} focused on the paramagnetic to
ferrimagnetic transition, and a very good agreement is found. It should be
emphasized that those authors have refined the patterns using the expanded
``222" cell for the nuclear phase in the ferrimagnetic region in the $Pmma$
space group, an hypothesis which ---as we mentioned before--- we do not
discard but prefer to use the ``122" $Pmmm$ cell to be consistent along the
whole series. For the lowest temperature, however, those authors did not
present any refinement of their data. The intriguing fact that the ``222"
magnetic reflections reappeared at 190~K after having disappeared at 265~K was
left unexplained. In the present work, we show that our diffractogram at 70~K
can be satisfactorily refined in the framework of the analysis presented for
the $x_{\rm{Ca}}=0.05$ sample. Therefore, the nuclear diffraction was
accounted for by using two orthorhombic phases, \emph{O}1 and \emph{O}2, and
the whole set of magnetic reflections could then be assigned to each of these
phases. As in $x_{\rm{Ca}}=0.05$, the \emph{O}2 phase presents an AFM ordering
with a ``224" magnetic cell. The \emph{O}1 phase, on the other hand, cannot
keep its ferrimagnetic SSO order because this would not be compatible with the
macroscopic magnetization measurements. In addition, a close inspection of the
magnetic peaks reveals that not all the reflections from the 273~K
ferrimagnetic phase are present, but that contributions to intensity related
to FM planes are absent at 70~K. Therefore, we have used a second AFM model
for the \emph{O}1 phase at 70~K, with a ``222" magnetic cell, and a
G--type--like ordering, although we have allowed for different magnetic moment
values in pyramids and octahedra. The model is schematized in
Fig.~\ref{f:estructura}(d). This scenario is also supported by our data on a
different sample, substituted with 5~\% Sr (
Y(Ba$_{0.95}$Sr$_{0.05}$)Co$_2$O$_{5.5}$) for which we have also performed
neutron thermodiffraction scans and apparently behaves almost the same as the
parent compound. These results will be published in a separate
paper.~\cite{inprogress}

\begin{table}[bt]
\caption{Structural parameters refined from the high resolution D2B data for
the parent compound YBaCo$_2$O$_{5.5}$ at $T=70$~K, 273~K and 348~K . Two sets
of atomic fractional coordinates are given for the \emph{O}1 and \emph{O}2
phases, which correspond to space group $Pmmm$ in the following Wyckoff
positions: Y (\emph{2p})=($\frac{1}{2},y,\frac{1}{2}$); Ba, Ca
(\emph{2o})=($\frac{1}{2}, y, 0)$; CoOct
(\emph{2r})=($0,\frac{1}{2},\frac{1}{4}$); CoPyr (\emph{2q})= ($0,0,z$); O1
(\emph{1a})=($0,0,0$); O2 (\emph{1e})=($0,\frac{1}{2},0$); O3
(\emph{1g})=($0,\frac{1}{2},\frac{1}{2}$); O3'
(\emph{1c})=($0,0,\frac{1}{2}$); O4 (\emph{2s})=($\frac{1}{2},0,z$); O5
(\emph{2t})=($\frac{1}{2},\frac{1}{2},z$); O6 (\emph{4u})=($0,y,z$).}
\label{t:VIa}
\begin{ruledtabular}
\begin{tabular}{lccccc}
 & & \multicolumn{2} {c} {$T=70$~K}& $T=273$~K & $T=348$~K \\
 &  &  \emph{O}1 & \emph{O}2 & \emph{O}1 & \emph{O}1\\ \hline
Y & $y$ & 0.2809(9)  & 0.271(1)   & 0.2745(6) & 0.2692(5) \\
Ba & $y$ & 0.255(2)  & 0.238(2)  & 0.254(1) & 0.244(1) \\
CoP & $z$ & 0.251(3)  & 0.270(3)  & 0.261(1) & 0.260(1) \\
O4 & $z$ & 0.321(2)  & 0.300(2)  & 0.3127(9) & 0.3126(7) \\
O5 & $z$ & 0.251(2)  & 0.276(3)  & 0.274(1) & 0.267(1) \\
O6 & $y$ & 0.2449(9)  &  0.243(1) & 0.2495(9) & 0.2437(8) \\
O6 & $z$ & 0.299(1)  & 0.300(1)  & 0.2935(7) & 0.2984(5) \\
O3 & $Occ$ &  1.0 & 0.67(6)  & 0.86(3) & 0.81(2)  \\
O3' & $Occ$ & 0.0  & 0.18(4)  & 0.05(2) & 0.12(2) \\
$a$ ({\AA}) & & 3.8515(2) & 3.8819(2) & 3.8496(1) & 3.8221(1) \\
$b$ ({\AA}) & & 7.7785(5) & 7.7156(4) & 7.8085(2) & 7.8581(3) \\
$c$ ({\AA}) & & 7.4859(6) & 7.4845(6) & 7.5032(2) & 7.5250(2) \\
$f$ (\%) & & 42(3) & 58(3) & 100 & 100\\
$R_{B}$ &  & 7.7  & 7.5  & 8.1 & 6.4 \\
$\chi^2$ & & \multicolumn{2} {c} {3.98} & 3.97 & 3.53 \\
\end{tabular}
\end{ruledtabular}
\end{table}

In Table~\ref{t:VIa} we present the details of the refined structures for the
sample with $x_{\rm{Ca}}=0$ from D2B data. The diffractogram at 70~K was
refined with the \emph{O}1+\emph{O}2 mixture, plus two AFM phases associated
to each of them, one having a ``224" supercell and the other one with a ``222"
supercell. At 230~K, only the \emph{O}1 phase with ferrimagnetic order was
enough to refine the diffractogram, whereas at 348~K just the nuclear
\emph{O}1 phase was refined. It is worth noting that the parent compound seems
to have a higher degree of misplaced octahedra when compared to the
Ca--substituted samples. This is evidenced by the non--zero occupation of the
O3' site, which corresponds to the empty apical oxygen position of pyramids.
At the highest temperature, however, there is a slight rearrangement of
vacancies among the O3 and O3' sites. When comparing the low temperature
phases \emph{O}1 and \emph{O}2 in the $x_{\rm{Ca}}=0$ and $x_{\rm{Ca}}=0.05$
samples, we observe that the \emph{O}1 phase seems to prefer a more perfect
order of pyramids and octahedra, while the excess oxygen vacancies accommodate
in the \emph{O}2 phase. There is also consistency among the structural
parameters in both samples for the \emph{O}1 and \emph{O}2 phases. The results
for the refined lattice parameters at the three temperatures studied are shown
in Fig.~\ref{f:LPs}(c). Our data are compared with those published by Akahoshi
and Ueda (diamonds).~\cite{99Aka} The difference at 70~K is due to the use of
one or two phases to refine the data. We have found no way of indexing the
whole set of magnetic reflections based on a single nuclear structure. It has
been shown in other \emph{R} cobaltites that there could be two coexisting
magnetic arrangements on a single nuclear structure,~\cite{02Fau,06Fro}
however, the evidence found in the $x_{\rm{Ca}}=0.05$ thermodiffractograms,
and the consistency obtained in the whole series when adopting such a phase
separation model, give us confidence in the proposed scenario. The complexity
of the systems seems to be related to the small size of the \emph{R} cation.

\section{Concluding remarks}

Although the layered cobaltites \emph{R}BaCo$_{2}$O$_{5.5}$ have received
great attention in the past five years, much of its behavior remains still
controversial and unclear. In this paper, an attempt is made to get some
insight into the role of cationic disorder by substituting the Ba--site, a
topic that has not yet been investigated to the best of our knowledge.
Interestingly, the systematics of this substitution led us to clarify and to
propose a model that describes the behavior at low temperature of the undoped
parent compound. Even though Ca addition does not lead to severe structural
distortions, it has nevertheless dramatic effects on the magnetic arrangement
and stability of the ferrimagnetic phase on detriment of the AFM long--range
order. Our results open up the possibility of studying Ca--doped cobaltites in
order to isolate the intrinsic properties of the ``122" ferrimagnetic phase in
monophasic samples, avoiding spurious effects in the analysis of macroscopic
properties. Further work is in progress to investigate the role of different
cations substitution and the systematics of the Ba--site disorder effects.


\begin{acknowledgments}

This work is part of a research project supported by Agencia Nacional de
Promoci\'{o}n Cient\'{\i }fica y Tecnol\'{o}gica (Argentina), under grant PICT
17-21372 and 20144, by CONICET (Argentina) under grant PIP 5250/05 and
5657/05, and by SECTyP, Universidad Nacional de Cuyo. JC acknowledges a
fellowship from CNEA and CONICET. We particularly acknowledge ILL and its
staff for the beamtime allocation and technical assistance.

\end{acknowledgments}



\end{document}